# Black Phosphorus-Monolayer MoS$_2$ van der Waals Heterojunction P-N Diode

Yexin Deng[†,*], Zhe Luo[§], Nathan J. Conrad[†], Han Liu[†], Yongji Gong[‡],

Sina Najmaei[‡], Pulickel M. Ajayan[‡], Jun Lou[‡], Xianfan Xu[§,*], Peide D. Ye[†,*]

[†] School of Electrical and Computer Engineering and Birck Nanotechnology Center, Purdue University, West Lafayette, IN 47907, USA

[§] School of Mechanical Engineering and Birck Nanotechnology Center, Purdue University, West Lafayette, IN 47907, USA

[‡] Department of Mechanical Engineering and Materials Science, Rice University, Houston, TX 77005, USA

Address correspondence to:

*(Y. Deng) deng58@purdue.edu, *(X. Xu) xxu@purdue.edu, *(P. D. Ye) yep@purdue.edu

# Abstract

Phosphorene, an elemental 2D material, which is the monolayer of black phosphorus, has been mechanically exfoliated recently. In its bulk form, black phosphorus shows high carrier mobility (~10000 cm$^2$/V s) and a ~0.3 eV direct bandgap. Well-behaved p-type field-effect transistors with mobilities of up to 1000 cm$^2$/V s, as well as phototransistors, have been demonstrated on few-layer black phosphorus, showing its promise for electronics and optoelectronics applications due to its high hole mobility and thickness-dependence direct bandgap. However, p-n junctions, the basic building blocks of modern electronic and optoelectronic devices, have not yet been realized based on black phosphorus. In this paper, we demonstrate a gate tunable p-n diode based on a p-type black phosphorus/n-type monolayer MoS$_2$ van der Waals p-n heterojunction. Upon illumination, these ultra-thin p-n diodes show a maximum photodetection responsivity of 418 mA/W at the wavelength of 633 nm, and photovoltaic energy conversion with an external quantum efficiency of 0.3%. These p-n diodes show promise for broadband photodetection and solar energy harvesting.

The successful isolation of graphene from graphite has led to its extensive study in physics, materials, and nano-engineering due to its extraordinary electrical and mechanical properties.[1-4] However, a lack of a bandgap limits its potential for electronic device applications, and has inspired the exploration of other 2D layered materials.[5-7] Among them, transition metal dichalcogenides (TMDCs), such as $MoS_2$, are the most studied materials.[8-11] Recently, phosphorene, the monolayer form of black phosphorus, has been successfully isolated.[12] Analogous to graphite and graphene, black phosphorus is a stack of phosphorene monolayers, bound together by van der Waals interactions.[12,13] Bulk black phosphorus shows a ~0.3 eV direct bandgap and a mobility of up to ~10000 $cm^2$/V s.[14-17] Its bandgap increases as its thickness decreases, and is predicted to have a >1 eV direct bandgap in its monolayer form.[12,13] Well-behaved p-type field-effect transistors with mobilities of up to 1000 $cm^2$/V s, as well as inverters, have been demonstrated on few-layer black phosphorus.[12,13,18-20] Based on its direct bandgap, few-layer black phosphorus phototransistors have been demonstrated with a responsivity of 4.8 mA/W.[19] These results indicate that black phosphorus is a promising candidate for both high performance electronics and optoelectronics applications due to its ultra-thin 2D nature, high hole mobility, and narrower direct bandgap compared to most of TMDCs.

P-N junctions are the basic building blocks of modern semiconductor devices, including diodes, bipolar transistors, photodiodes, light-emitting diodes, and solar cells. In the conventional p-n homo-junction, the p- and n-type regions are formed by chemically doping a bulk semiconductor, creating a graded junction region. P-N heterojunctions can be realized by epitaxially growing an n-type semiconductor on another p-type semiconductor or vice versa, and can form abrupt p-n junctions. Relying on the van der Waals interactions, atomically sharp 2D heterostructures can be achieved without the problems caused by lattice mismatches between materials. Moreover, the ultra-thin nature of 2D heterostructures allows for electrical modulation of the band structure in the vertical direction. This creates a new avenue for the realization of novel 2D electronic and optoelectronic devices.[21-33]

Here, we report an electrically tunable black phosphorus-monolayer $MoS_2$ van der Waals heterojunction p-n diode. To the best of our knowledge, this is the first 2D heterostructure demonstrated using black phosphorus. The p-type black phosphorus and n-type monolayer $MoS_2$ form an atomically sharp type II hetero-interface through van der Waals interactions.[13,34] The p-n diode exhibits gate tunable current-rectifying characteristics. Upon illumination, the p-n diode can be used as photodetector with a maximum responsivity of 418 mA/W, which is nearly 100 times higher than that reported for few-layer black phosphorus phototransistors, and 26 times higher than that recently reported for $WSe_2$ p-n diodes.[19,35,36] The photovoltaic power generation in the diode reaches a peak external quantum efficiency of ~0.3%. Furthermore, photocurrent mapping confirms that the current is generated throughout the entire overlapped p-n junction region, showing its feasibility for use in large area solar cells and photodetectors.

## Results and Discussion

In the experiments, monolayer $MoS_2$ was synthesized by chemical vapor deposition (CVD) on a 285 nm $SiO_2$/p+ doped Si substrate using $MoO_3$ and sulfur. More information about the synthesis method can be found in our previous papers.[37,38] Few-layer black phosphorus was mechanically exfoliated using adhesive tape from bulk material onto a $SiO_2$ substrate on which CVD monolayer $MoS_2$ had been synthesized. Due to van der Waals interactions, 2D black phosphorus-monolayer $MoS_2$ heterojunctions can be formed at the overlapped regions. Electron-beam lithography was used to define the contact patterns. Finally, 20/60 nm of Ni/Au were deposited as metal contacts. Figure 1a, b show the schematics of the device structure and the electrical connections. A voltage ($V_d$) was applied across the diode, and a voltage ($V_g$) was applied to the p+ doped silicon as a back gate, which can electrically modulate the device. The optical image of the fabricated device is shown in Figure 1c, and all the following results were obtained from this particular device unless otherwise specified. To achieve better current collection and reduce series resistance, the $MoS_2$ electrode is designed to surround the black phosphorus flake.

The thicknesses of the few-layer black phosphorus flake and monolayer $MoS_2$ are ~11 nm and ~0.9 nm, respectively, as determined from the atomic force microscopy measurements shown in Fig. 2a. The monolayer nature of the $MoS_2$ can be further confirmed by the peak position (~1.8 eV) in its photoluminescence spectrum, shown in Figure 2b.[39] The Raman spectra from the black phosphorus, monolayer $MoS_2$ and the heterojunction regions are presented in Figure 2c. The observed Raman-active modes of black phosphorus and $MoS_2$ are consistent with previously reported data [12,37] The peaks of both $MoS_2$ and black phosphorus can be observed in the overlapped region with very little shift from their positions in the non-overlapped regions, indicating good quality of thin films in the junction region after exfoliation and device fabrication.

Next, we studied the electrical characteristics of the fabricated device. In this paper, all the measurements were performed in ambient atmosphere. Figure 3a shows the gate tunable I-V characteristics of the p-n diode, and the inset (1) shows the I-V curves on a semi-log plot. The current-rectifying characteristics can be modulated by the back gate voltage, shown in the inset (2) of Figure 3a. The rectification ratio, defined as the ratio of the forward/reverse current, increases as the back gate voltage decreases. By using -30 V back gate voltage, a rectification ratio of ~$10^5$ is obtained at $V_d$ = -2/+2 V. Additionally, the ideal factor of the p-n diode increases as the back gate voltage increases (Figure S8), and achieve a minimum value of 2.7 with a back gate voltage of -30V. These strong current-rectifying characteristics indicate that a good van der Waals p-n heterojunction formed between p-type black phosphorus and n-type $MoS_2$. The modulation effect of the back gate voltage can also be seen in the transfer curves (Figure 3b). Both forward and reverse currents can be substantially increased by increasing the back gate voltage. These results can be explained by a simplified model describing the current transport of the p-n diode (See SI for more details). For simplicity, the total resistance of the device can be roughly divided into three parts: the resistance of the p-n junction near the interface, the sheet resistances of $MoS_2$ and few-layer black phosphorus, and the contact resistances of metal/$MoS_2$ and metal/black phosphorus. The current-rectifying characteristics come from the p-n interface region, which may be modeled as many parallel p-n diodes (Figure S2a). The band alignment of $MoS_2$ and black phosphorus at the p-n junction interface can be modulated by the back gate voltage, as shown in Figure S3. This can explain why the rectification ratio increases as the back gate voltage decreases, and the ideal factor decreases as the back gate voltage decreases. Moreover, the back gate voltage can also modulate the sheet resistance and the contact resistance. As the device size is relatively large, the sheet resistances are more significant than the contact resistances. Because the sheet resistance of CVD monolayer $MoS_2$ is much larger than the 11 nm black phosphorus, the overall device resistance is strongly impacted by the resistance of $MoS_2$.[12,40] By increasing the back gate voltage ($V_g$), both the resistance of $MoS_2$ and its contact resistance are reduced, which can efficiently boost the total current under both forward and reverse bias (See the SI for more details).

As good current-rectifying characteristics were achieved, the optoelectronic characteristics of these 2D p-n diodes were then explored. A 633 nm He-Ne laser was used to illuminate the device. The spot size of the laser was controlled to be much smaller than the junction area in order to exclude the photoresponse from the non-overlapped region. Figure 4a shows the IV curves of the p-n diode under various incident laser power, and the inset shows the details of the negative $V_d$ bias region. The photocurrent $I_{ph}$ is defined as $I_{illumination}$ - $I_{dark}$ where $I_{illumination}$ and $I_{dark}$ are the $I_d$ with and without laser illumination. Under reverse bias, the photocurrent has a strong dependence on the incident power (Figure 4b). By increasing the back gate voltage, the photocurrent is substantially increased. This can be attributed to the reduction of the $MoS_2$ sheet resistance and the $MoS_2$/metal contact resistance (see SI for more information). On the other hand, by applying a strong negative back gate voltage (-40 V) to modulate the band alignment of $MoS_2$ and black phosphorus (Figure S3), the ratio of the illumination current over the dark current ($I_{illumination}/I_{dark}$) can be increased to $3 \times 10^3$, though the illumination current decreases due to the increasing of sheet and contact resistance (see SI for more details). This is ~100 times larger than that of a $MoS_2$ photodetector due to a better suppression of $I_{dark}$ by the reversed bias p-n diode. Compared with the phototransistor, it is beneficial for the device to detect signal from noise.[41] With these results, the photodetection responsivity R, defined as $I_{ph}/P_{laser}$ where $P_{laser}$ is the incident power of the laser, can be calculated for both forward (as photoconductor) and reverse bias (as photodiode) as shown in Figure 4c. The responsivity of both the forward and the reverse bias regions decrease as the laser power increases, which is due to the saturation of electron-hole pair generation at higher incident light intensity and the increased surface recombination in either the junction region or the underlying $SiO_2$.[41] Therefore, to further improve the responsivity, it is necessary to improve the film quality of the CVD MoS$_2$ and black phosphorus, as well as the dielectric interface quality. In this device, the responsivity reaches 1.27 A/W and 11 mA/W for forward and reverse bias, respectively. The maximum responsivity in our devices is up to 3.54A/W (at V$_d$ = +2V) and 418 mA/W (at V$_d$ = -2V) under 1μW laser power (Figure S5). As a photodiode, this is nearly 100 times higher than that of the recently reported black phosphorus phototransistor, and 4.8 times higher than that of the carbon nanotube-MoS$_2$ p-n diode with a much smaller V$_d$.[19,32] By using this electrically tunable 2D p-n diode, it is possible to realize sensitive and broadband photodetection due to the high mobility and relatively small direct bandgap of few-layer black phosphorus.

In addition to the photodetection, this p-n diode is also usable for photovoltaic energy conversion. Under laser illumination, the short circuit current (I$_{sc}$), which is the current through the device with zero volts bias, is shown in Figure 5a. The open circuit voltage (V$_{oc}$) can be obtained by examining the V$_d$-axis intercepts in the plot. As seen in Figure 5c, d, both the I$_{sc}$ and V$_{oc}$ increase as the laser power increases. However, the back gate modulates the I$_{sc}$ and V$_{oc}$ in different ways. Increasing the back gate voltage boosts the I$_{sc}$ while reducing the V$_{oc}$. This can be attributed to the opposite modulation effect of band alignment of MoS$_2$/black phosphorus at the junction interface, and the sheet and contact resistance of MoS$_2$ (see SI for details). With this data, the power generated by the p-n diode can be calculated by P$_d$ = I$_d$V$_d$, as shown in Figure 5b. Then, the fill factor, which is defined as FF=P$_d$/(I$_{sc}$V$_{oc}$), can be obtained. The maximum FF is ~0.5 in this 2D p-n diode. Furthermore, the external quantum efficiency (EQE) can be calculated by EQE = ($I_{sc}/P_{laser}$) (hc/e$\lambda$), where h is the Planck constant, e is the electron charge, c is the velocity of light, $\lambda$ is the wavelength of the incident light. The peak EQE in this device is 0.3%. To the best of our knowledge, this is the first time that efficient photovoltaic energy conversion has been demonstrated using black phosphorus. By further reducing the thickness of the black phosphorus to bilayer, it is possible to essentially enhance the EQE to up to 18%, according to theoretical predictions.[42] Moreover, by stacking the 2D p-n diode vertically, a stacked solar cell structure can be realized. Since the direct bandgap of black phosphorus changes with its thicknesses, the efficiency may be further improved by tuning its thickness to better utilize the photon energies of different wavelengths of the solar spectrum.[12]

To further understand the photoresponse process, photocurrent mapping was performed in order to determine the spatial distribution of the photocurrent generation. A device with a larger junction region was chosen to get more spatial information, as shown in Figure 6a. The thickness of the black phosphorus flake is ~22 nm (Figure S6). A 633 nm He-Ne laser was used with ~8 μW power and ~1 μm spot size. The current was measured under zero volts bias. As seen in the Figure 6b, the current is generated throughout the junction region, which indicates the charge separation mainly in the heterojunction region. However, the photoresponse across the junction is not uniform. This can be attributed to the non-uniform contacts between the two flakes. Moreover, the part of the junction region that near the metal contact on the MoS$_2$ exhibits stronger photocurrent. This is related to the lateral sheet resistance of MoS$_2$ and black phosphorus. Under illumination, the charges are separated near the vertical interface p-n diode region. For simplicity, we treat the interface p-n diode (see Figure S4) at a certain position as a voltage source under illumination. Here we assume the value of this voltage source is uniform across the junction, if different positions are under laser illumination. The photocurrent flows though MoS$_2$ and black phosphorus to reach the metal contact (Figure S4). Because the sheet resistance of CVD monolayer MoS$_2$ is much larger than the black phosphorus, the resistance of the MoS$_2$ makes a much stronger impact. The part of the junction region closer to the metal contact (B in Figure S4, compared with A) on MoS$_2$ shows a lower MoS$_2$ sheet resistance, resulting in the stronger photocurrent near the contact of MoS$_2$ (see SI for details). To get a more uniform spatial photoresponse, a transparent contact material, such as graphene, could be used as top electrodes. This can increase the contact area and yield better carrier collection. Furthermore, graphene can also reduce the contact resistance as an interlayer between the metals and the TMDCs, further improving the device efficiency.[43] This scheme makes it possible for the black phosphorus/MoS$_2$ heterojunction to be used as large area transparent and flexible solar cells, as well as photodetectors.

## Conclusion

In conclusion, for the first time, we demonstrate a gate tunable black phosphorus-monolayer $MoS_2$ van der Waals heterojunction p-n diode. These 2D p-n diodes exhibit strong gate tunable current-rectifying IV characteristics, which is investigated using a simplified device model. As a photodetector, the p-n diode shows a photodetection responsivity of 418 mA/W which is nearly 100 times higher than the reported black phosphorus phototransistor. Due to the small bandgap and high mobility of few-layer black phosphorus, it is very suitable for broadband and sensitive photodetection. Our p-n diodes also show photovoltaic power generation with a peak external quantum efficiency of 0.3%, which can be further improved by reducing the thickness of black phosphorus. Furthermore, photocurrent mapping is performed to study the spatial distribution of photoresponse, which reveals the importance of large area transparent electrodes for 2D p-n diodes to achieve large area optoelectronics applications.

Methods:

First, monolayer $MoS_2$ was synthesized using chemical vapor deposition method on heavily doped silicon wafer capping with 285 nm $SiO_2$. The black phosphorus was mechanically exfoliated from the bulk black phosphorus onto the wafer. The black phosphorus-monolayer $MoS_2$ heterojunction was achieved through the van der Waals interactions. Then the standard e-beam lithography process was used to define contact patterns and pads. 20/60 nm Ni/Au was deposited as metal contacts. The electrical measurements were performed using Keithley 4200 Semiconductor Characterization System. All the electrical and optical measurements were performed in the ambient atmosphere.

*Conflict of interests*: The authors declare no competing financial interest.

*Acknowledgements*: We thank Prof. Jong-Hyun Choi, Haorong Chen and Te-Wei Wen from Purdue University for help with AFM characterization. This material is based upon work partly supported by NSF under Grant CMMI-1120577 and SRC under Task 2396.

*Supporting Information Available*: IV characteristics of various devices, analysis of current transport in these p-n didoes, photodetection responsivity of a p-n diode, and thickness data of the few-layer black phosphorus flake for photocurrent mapping measurements. This material is available free of charge *via* the Internet at http://pubs.acs.org

References and Notes:

(1) Geim, A. K.; Novoselov, K. S. The Rise of Graphene. *Nat. Mater.* **2007**, *6*, 183–191.

(2) Zhang, Y.; Tan, Y.-W.; Stormer, H. L.; Kim, P. Experimental Observation of the Quantum Hall Effect and Berry's Phase in Graphene. *Nature* **2005**, *438*, 201–204.

(3) Geim, A. K. Graphene: Status and Prospects. *Science* **2009**, *324*, 1530–1534.

(4) Novoselov, K. S.; Fal'ko, V. I.; Colombo, L.; Gellert, P. R.; Schwab, M. G.; Kim, K. A Roadmap for Graphene. *Nature* **2012**, *490*, 192–200.

(5) Novoselov, K.; Jiang, D.; Schedin, F.; Booth, T.; Khotkevich, V.; Morozov, S.; Geim, A., Two-dimensional Atomic Crystals. *Proc. Natl. Acad. Sci. U.S.A.* **2005**, *102*, 12451-10453.

(6) Xu, M.; Liang, T.; Shi, M.; Chen, H. Graphene-like Two-dimensional Materials. *Chem. Rev.* **2013**, *113*, 3766–3798.

(7) Butler, S.; Hollen, S.; Cao, L.; Cui, Y.; Gupta, J.; Gutierrez, H.; Heinz, T.; Hong, S.; Huang, K.; Ismach, A. *et al*, Progress, Challenges, and Opportunities in Two-dimensional Materials beyond Graphene. *ACS Nano* **2013**, *7*, 2898–2926.

(8) Wang, Q. H.; Kalantar-Zadeh, K.; Kis, A.; Coleman, J. N.; Strano, M. S. Electronics and Optoelectronics of Two-dimensional Tansition Metal Dichalcogenides. *Nature Nanotechnol*, **2012**, *7*, 699–712.

(9) Jariwala, D.; Sangwan, V. K.; Lauhon, L. J.; Marks, T. J.; Hersam, M. C. Emerging Device Applications for Semiconducting Two-Dimensional Transition Metal Dichalcogenides. *ACS Nano,* **2014**, *8*, 1102-1120.

(10) Radisavljevic, B.; Radenovic, A.; Brivio, J.; Giacometti, V.; Kis, A. Single-layer $MoS_2$ Transistors. *Nat. Nanotechnol.* **2011,** *6*, 147–150.

(11) Liu, H.; Neal, A. T.; Ye, P. D. Channel Length Scaling of $MoS_2$ MOSFETs. *ACS Nano* **2012,** *6*, 8563–8569.

(12) Liu, H.; Neal, A. T.; Zhu, Z.; Luo, Z.; Xu, X.; Tománek, D.; Ye, P. D. Phosphorene: An Unexplored 2D Semiconductor with a High Hole Mobility. *ACS Nano* **2014**, *8*, 4033-4041.

(13) Li, L.; Yu, Y.; Ye, G. J.; Ge, Q.; Ou, X.; Wu, H.; Feng, D.; Chen, X. H.; Zhang, Y. Black Phosphorus Field-effect Transistors. *Nat. Nanotechnol.* **2014**, *9*, 372–377.

(14) Morita, A. Semiconducting Black Phosphorus. *Appl. Phys. A Solids Surfaces* **1986**, *39*, 227–242.

(15) Warschauer, D. Electrical and Optical Properties of Crystalline Black Phosphorus. *J. Appl. Phys.* **1963**, *34*, 1853–1860.

(16) Bridgman, P. M. Two New Modifications of Phosphorus. *J. Am. Chem. Soc.* **1914,** 36, 1344–1363.

(17) Keyes, R. W. The electrical properties of black phosphorus. *Phys. Rev.* **1953**, *92*, 580-584.

(18) Xia, F.; Wang, H.; Jia, Y. R. Rediscovering Black Phosphorus: A Unique Anisotropic 2D Material for Optoelectronics and Electronics. arXiv:1402.0270.

(19) Buscema, M.; Groenendijk, D. J.; Blanter, S. I.; Steele, G. A.; Der, H. S. J. Van; Castellanos-gomez, A. Fast and Broadband Photoresponse of Few-Layer Black Phosphorus Field-effect Transistors. arXiv:1403.0565.

(20) Koenig, S.P.; Doganov, R.A.; Schmidt, H.; Neto, A.H.; Oezyilmaz, B. Electric Field Effect in Ultrathin Black Phosphorus. *Appl. Phys. Lett.* **2014**, *104*, 103106.

(21) Yu, W. J.; Li, Z.; Zhou, H.; Chen, Y.; Wang, Y.; Huang, Y.; Duan, X. Vertically Stacked Multi-heterostructures of Layered Materials for Logic Transistors and Complementary Inverters. *Nat. Mater.* **2013**, *12*, 246–252.

(22) Yang, H.; Heo, J.; Park, S.; Song, H. J.; Seo, D. H.; Byun, K.-E.; Kim, P.; Yoo, I.; Chung, H.-J.; Kim, K. Graphene Barristor, a Triode Device with a Gate-controlled Schottky Barrier. *Science* **2012**, *336*, 1140–1143.

(23) Britnell, L.; Gorbachev, R. V; Jalil, R.; Belle, B. D.; Schedin, F.; Mishchenko, A.; Georgiou, T.; Katsnelson, M. I.; Eaves, L.; Morozov, S. V; *et al.* Field-effect Tunneling Transistor Based on Vertical Graphene Heterostructures. *Science* **2012**,

*335*, 947–950.

(24) Britnell, L.; Ribeiro, R. M.; Eckmann, A.; Jalil, R.; Belle, B. D.; Mishchenko, A.; Kim, Y.-J.; Gorbachev, R. V; Georgiou, T.; Morozov, S. V; *et al.* Strong Light-matter Interactions in Heterostructures of Atomically Thin films. *Science* **2013**, *340*, 1311–1314.

(25) Yu, W. J.; Liu, Y.; Zhou, H.; Yin, A.; Li, Z.; Huang, Y.; Duan, X. Highly Efficient Gate-tunable Photocurrent Generation in Vertical Heterostructures of Layered Materials. *Nat. Nanotechnol.* **2013**. *8*, 952-958.

(26) Bertolazzi, S.; Krasnozhon, D.; Kis, A. Nonvolatile Memory Cells Based on MoS$_2$/Graphene Heterostructures. *ACS Nano* **2013**, 3246–3252.

(27) Georgiou, T.; Jalil, R.; Belle, B. D.; Britnell, L.; Gorbachev, R. V; Morozov, S. V; Kim, Y.-J.; Gholinia, A.; Haigh, S. J.; Makarovsky, O.; *et al.* Vertical Field-effect Transistor based on Graphene-WS$_2$ Heterostructures for Glexible and Transparent Electronics. *Nat. Nanotechnol.* **2013**, *8*, 100–103.

(28) Fang, H.; Battaglia, C.; Carraro, C.; Nemsak, S.; Ozdol, B.; Kang, J. S.; Bechtel, H. A.; Desai, S. B.; Kronast, F.; Unal, A. A.; *et al*. Strong Interlayer Coupling in van der Waals Heterostructures Built from Single-layer Chalcogenides. *Proc. Natl. Acad. Sci. U.S.A.* **2014.**

(29) Lee, C.; Lee, G.; Zande, A. M. Van Der; Chen, W.; Li, Y.; Han, M.; Cui, X.; Arefe, G.; Nuckolls, C.; Heinz, T. F.; *et al*. Atomically Thin P-N Junctions with van der Waals Heterointerfaces. arXiv:1403.3062.

(30) Furchi, M. M.; Pospischil, A.; Libisch, F.; Burgdörfer, J.; Mueller, T. Photovoltaic Effect in an Electrically Tunable van der Waals Heterojunction. arXiv:1403.2652.

(31) Cheng, R.; Li, D.; Zhou, H.; Wang, C.; Yin, A.; Jiang, S. el al, Electroluminescence and Photocurrent generation from Atomically Sharp $WSe_2/MoS_2$ Heterojunction P-N Diodes. arXiv:1403.3447.

(32) Jariwala, D.; Sangwan, V. K.; Wu, C.-C.; Prabhumirashi, P. L.; Geier, M. L.; Marks, T. J.; Lauhon, L. J.; Hersam, M. C. Gate-Tunable Carbon Nanotube-$MoS_2$ Heterojunction P-N Diode. *Proc. Natl. Acad. Sci. U.S.A*. **2013**, *110*, 18076-18080.

(33) Chuang, S.; Kapadia, R.; Fang, H.; Chia Chang, T.; Yen, W.-C.; Chueh, Y.-L.; Javey, A. Near-Ideal Electrical Properties of InAs/$WSe_2$ van Der Waals Heterojunction Diodes. *Appl. Phys. Lett.* **2013**, *102*, 242101.

(34) Lam, K.; Cao, X.; Guo, J. Device Performance of Heterojunction Tunneling Field-Effect Transistors Based on Transition Metal Dichalcogenide Monolayer. *IEEE Electron Device Lett.* **2013,** *34*, 1331–1333.

(35) Baugher, B. W. H.; Churchill, H. O. H.; Yang, Y.; Jarillo-Herrero, P. Optoelectronic Devices based on Electrically Tunable P-N Diodes in a Monolayer Dichalcogenide. *Nat. Nanotechnol.* **2014**, *9*, 262-267.

(36) Pospischil, A.; Furchi, M. M.; Mueller, T. Solar-energy Conversion and Light

Emission in an Atomic Monolayer P-N Diode *Nat. Nanotechnol.* **2014**, *9*, 257–261.

(37) Najmaei, S.; Liu, Z.; Zhou, W.; Zou, X.; Shi, G.; Lei, S.; Yakobson, B. I.; Idrobo, J.-C.; Ajayan, P. M.; Lou, J. Vapour Phase Growth and Grain Boundary Structure of Molybdenum Disulphide Atomic Layers. *Nat. Mater.* **2013**, *12*, 754–759.

(38) Liu, H.; Si, M.; Deng, Y.; Neal, A. T.; Du, Y.; Najmaei, S.; Ajayan, P. M.; Lou, J.; Ye, P. D. Switching Mechanism in Single-Layer Molybdenum Disulfide Transistors: An Insight into Current Flow across Schottky Barriers. *ACS Nano* **2013**, *8*, 1031–1038.

(39) Splendiani, A.; Sun, L.; Zhang, Y.; Li, T.; Kim, J.; Chim, C.-Y.; Galli, G.; Wang, F. Emerging Photoluminescence in Monolayer $MoS_2$. *Nano Lett.* **2010**, *10*, 1271–1275.

(40) Liu, H.; Si, M.; Najmaei, S.; Neal, A.; Du, Y.; Ajayan, P.; Lou, J.; Ye. P. D., Statistical Study of Deep Submicron Dual-Gated Field-Effect Transistors on Monolayer Chemical Vapor Deposition Molybdenum Disulfide Films. *Nano Lett.* **2013**, *13*, 2640–2646.

(41) Lopez-Sanchez, O.; Lembke, D.; Kayci, M.; Radenovic, A.; Kis, A. Ultrasensitive Photodetectors based on Monolayer $MoS_2$. *Nat. Nanotechnol.* **2013**, *8*, 497–501.

(42) Dai, J.; Zeng, X. Bilayer Phosphorene: Effect of Stacking Order on Bandgap and its Potential Application in Thin-Film Solar Cells. *J. Phys. Chem. Lett.* **2014**, 5,

1289–1293.

(43) Du, Y.; Yang, L.M.; Zhang, J.Y.; Liu, H.; Majumdar, K.; Kirsch, P.D.; Ye, P.D. MoS2 Field-Effect Transistors With Graphene/Metal Heterocontacts. *IEEE Electron Device Letters* **2014**, *5*, 599-601.

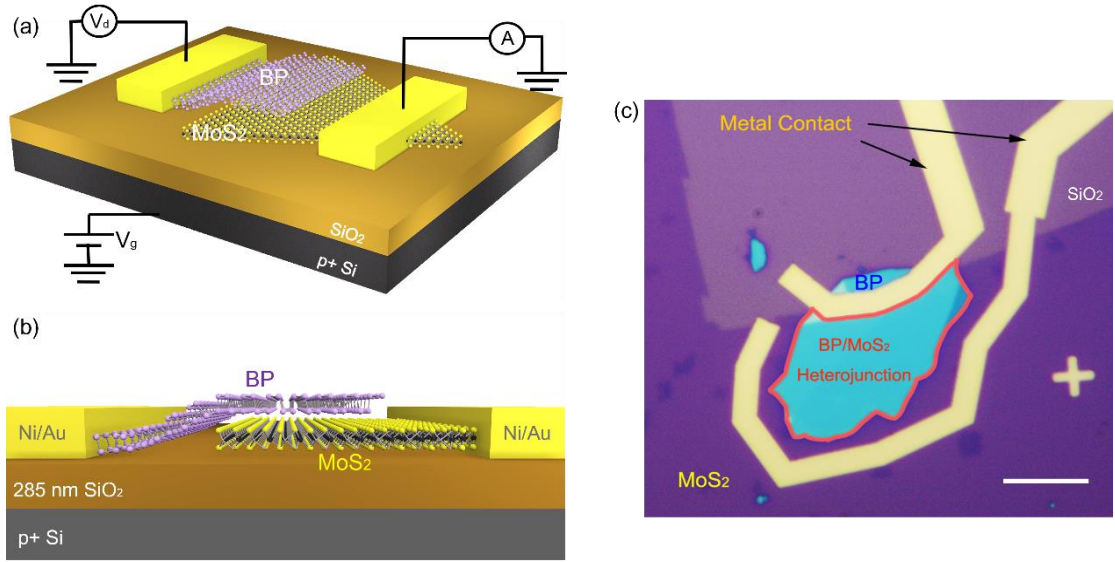

Figure 1. (a) and (b) Schematics of the device structure. A p+ silicon wafer capped with 285 nm $SiO_2$ is used as the global back gate and the gate dielectric. Few-layer black phosphorus flakes were exfoliated onto monolayer $MoS_2$ in order to form a van der Waals heterojunction. Ni/Au were deposited as contacts. During the electrical measurements, a voltage $V_d$ is applied across the device. The voltage bias $V_g$ is applied to the back gate. (c) Optical image of the fabricated device. The dark purple region is monolayer $MoS_2$, while the blue flake is few-layer black phosphorus. The light purple region is $SiO_2$. Scale bar, 10 μm.

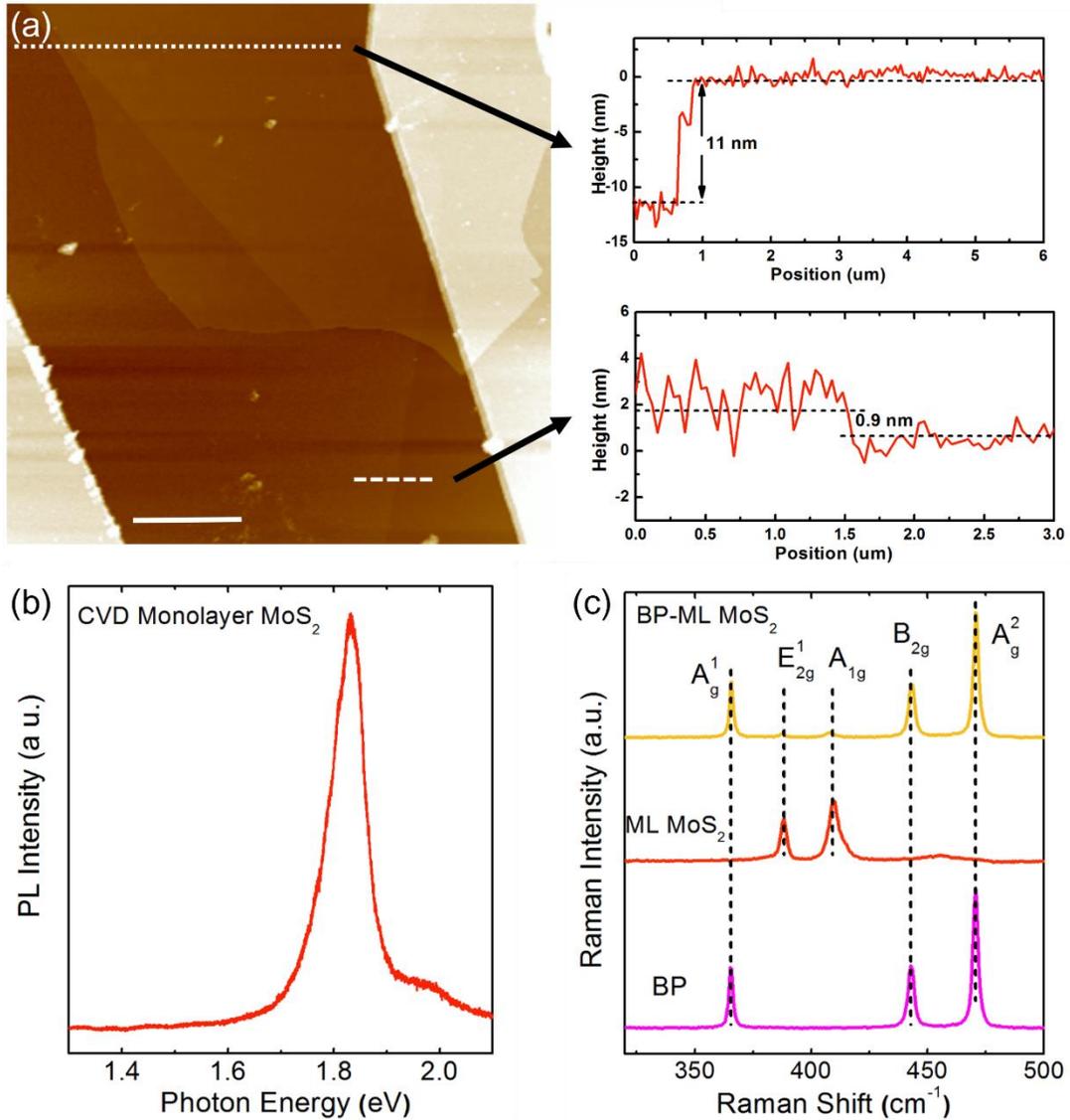

Figure 2. (a) Atomic force microscopy image of part of the junction region. It shows that the black phosphorus and MoS$_2$ films are ~11 nm and ~0.9 nm thick, respectively. Scale bar, 2 μm. (b) Photoluminescence spectrum of the MoS$_2$ thin film. It shows a strong peak close to 1.8 eV, confirming that the MoS$_2$ is a monolayer. (c) Raman spectra of the MoS$_2$, black phosphorus, and the overlapped regions. Both of the black phosphorus and MoS$_2$ Raman peaks can be observed in the overlapped region.

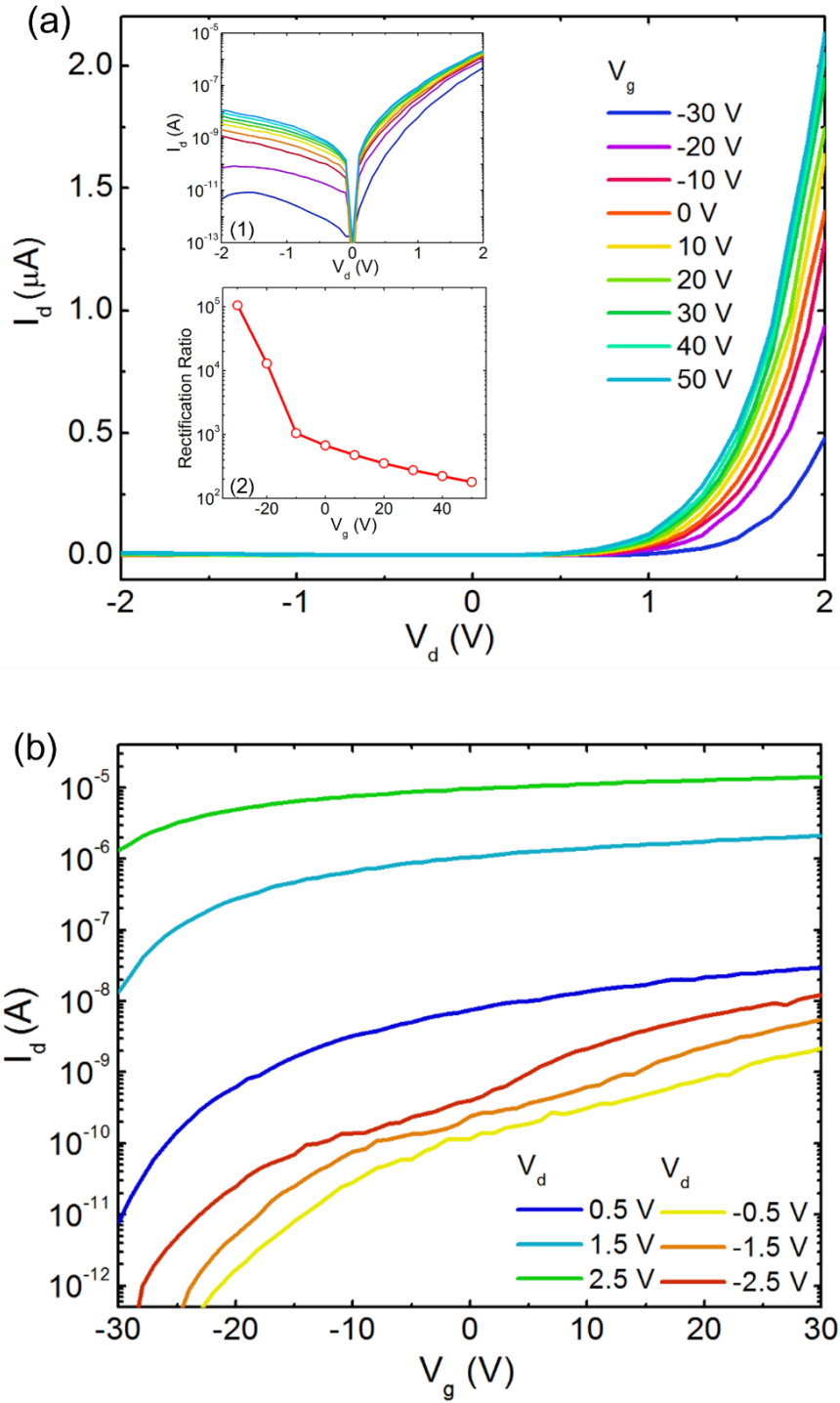

Figure 3. (a) Gate tunable IV characteristics of the 2D p-n diode. The current increases as the back gate voltage increases. The inset (1) shows the IV characteristics under semi-log scale. The inset (2) shows the rectification ratio as a function of back gate voltage $V_g$. (b) Transfer curves of p-n diode for both forward and reverse $V_d$ bias with back gate modulations.

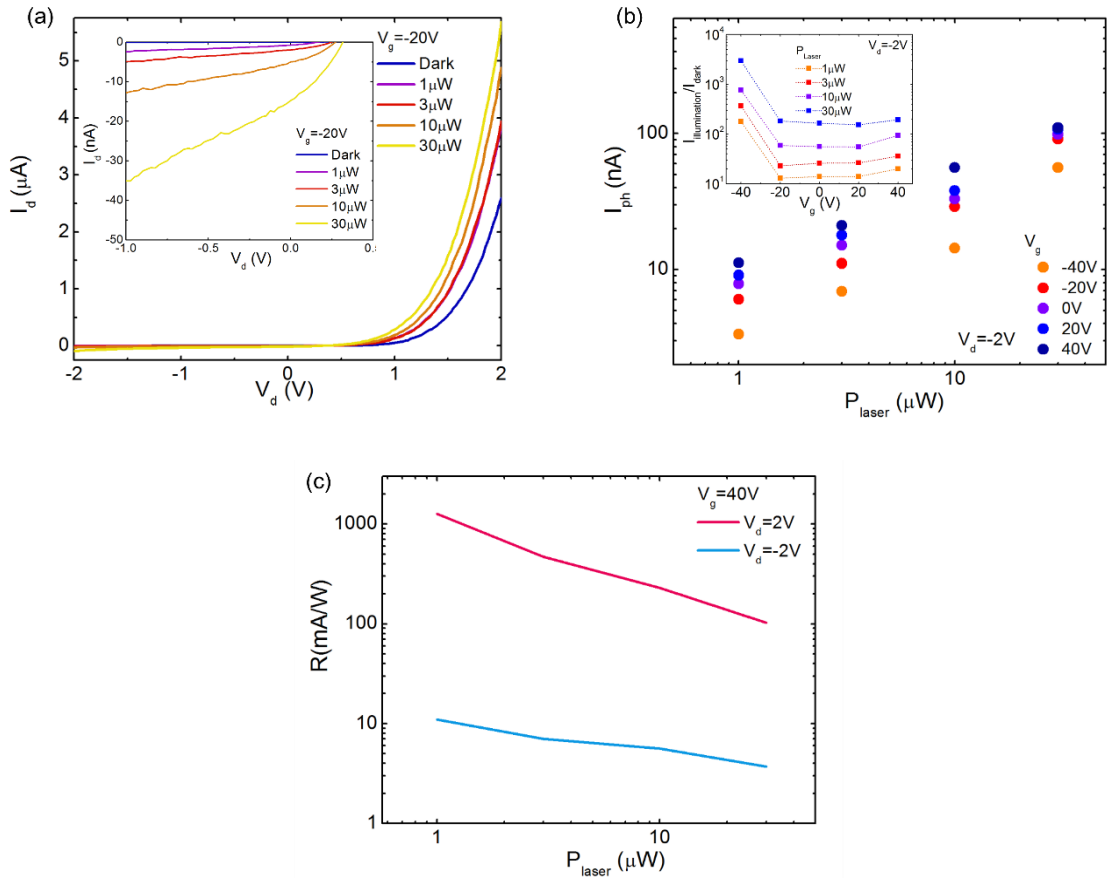

Figure 4. (a) IV characteristics of the p-n diode under various incident laser powers. The inset shows the details in the reverse bias region. (b) The photocurrent as a function of incident laser power. Increasing the back gate voltage can increase the photocurrent. The inset shows the ratio of $I_{illumination}/I_{dark}$. (c) Photodetection responsivity (R) calculated as a function of incident power. The responsivity decreases as the power increases.

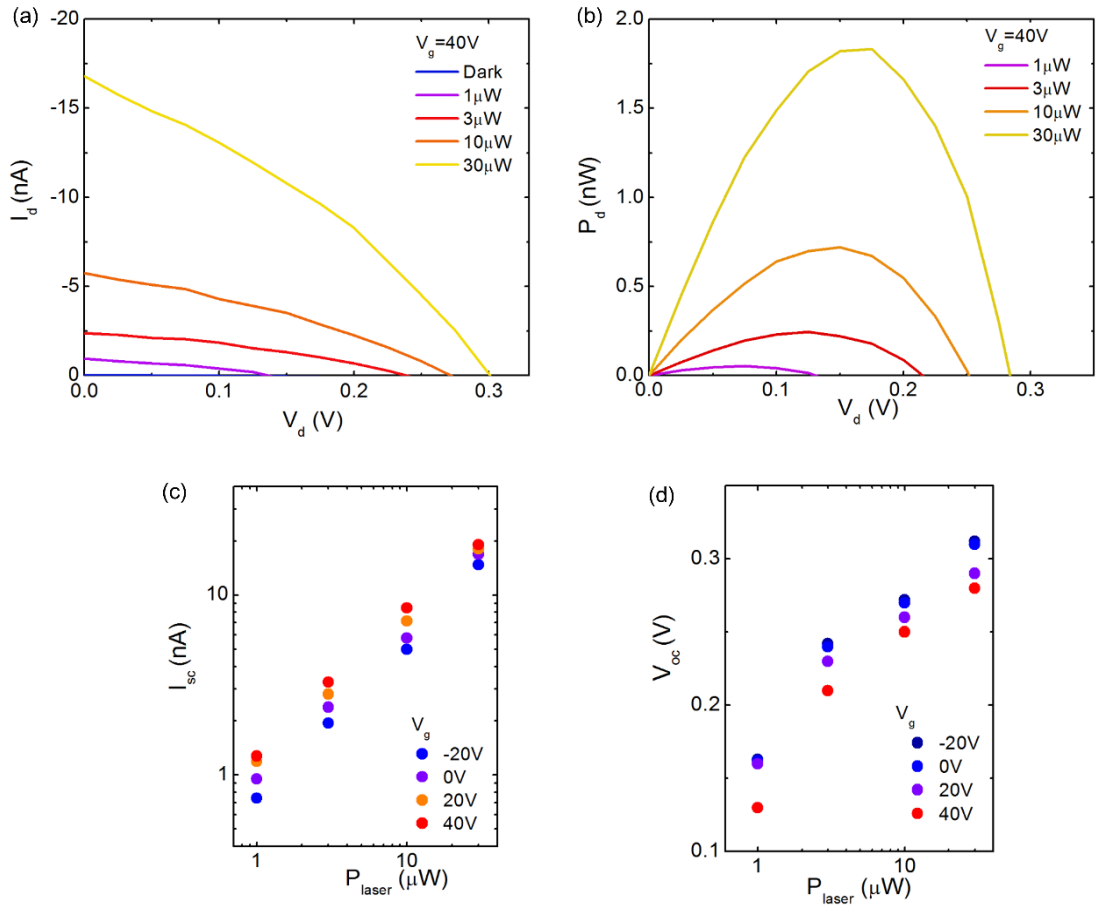

Figure 5. (a) $I_d$ as a function of function of $V_d$ under various laser powers. The $I_{sc}$ and $V_{oc}$ can be obtained from the intercepts of the curves on $I_d$ and $V_d$ axes. (b) Power generated by the p-n diode as a function of $V_d$ under different laser power. (c) $I_{sc}$ as a function of laser power under different back gate voltage. (d) $V_{oc}$ as a function of laser power under different back gate voltage. Increasing the back gate voltage increases $I_{sc}$, but reduces $V_{oc}$.

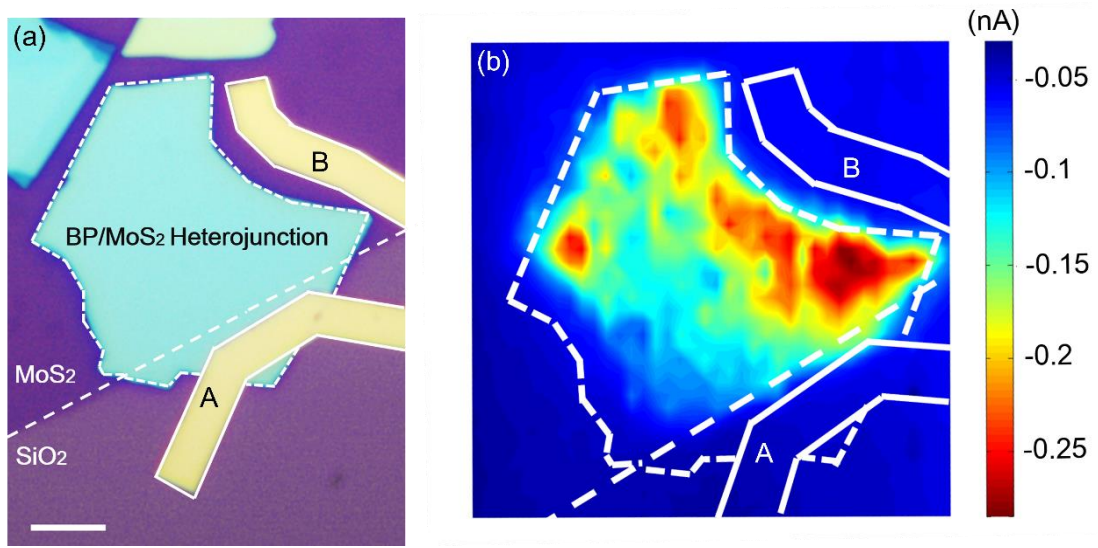

Figure 6. (a) Optical image of the p-n heterojunction used for photocurrent mapping. Scale bar, 5 μm. (b) Photocurrent mapping of the p-n heterojunction under zero volt bias. The dashed lines outline the heterojunction regions. The solid lines outline the metal contacts. Contact A and B are electrically connected to black phosphorus and $MoS_2$, respectively.

*Supporting Information*

# Black Phosphorus-Monolayer MoS$_2$ van der Waals Heterojunction P-N Diode

Yexin Deng[†,*], Zhe Luo[§], Nathan J. Conrad[†], Han Liu[†], Yongji Gong[‡],

Sina Najmaei[‡], Pulickel M. Ajayan[‡], Jun Lou[‡], Xianfan Xu[§,*], Peide D. Ye[†,*]

[†] School of Electrical and Computer Engineering and Birck Nanotechnology Center, Purdue University, West Lafayette, IN 47907, USA

[§] School of Mechanical Engineering and Birck Nanotechnology Center, Purdue University, West Lafayette, IN 47907, USA

[‡] Department of Mechanical Engineering and Materials Science, Rice University, Houston, TX 77005, USA

[*]Correspondence to: (Y. Deng) deng58@purdue.edu, (X. Xu) xxu@purdue.edu
(P. D. Ye) yep@purdue.edu.

**S1. IV characteristics of various p-n diodes**

**S2. Analysis of the current transport in the p-n diode**

**S3. Photodetection Responsivity of a p-n diode**

**S4. The thickness of the black phosphorus flake for the photocurrent mapping**

**S5. Analysis of ideal factor of the p-n diode**

## 1. IV characteristics of various p-n diodes

The characteristics of the 2D p-n diodes are reproducible in many devices. However, with different junction areas and contact qualities between the flakes, variations were observed. Figure S1a, b show the IV characteristics of two different devices.

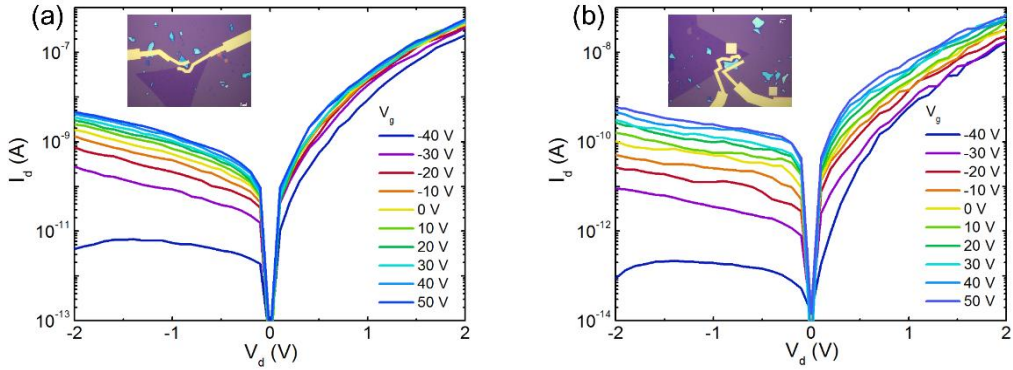

Figure S1 (a) and (b) show the gate tunable IV characteristics of two different p-n diodes.

## 2. Analysis of the current transport in the p-n diode

In the 2D p-n diode demonstrated in this work, the vertical van der Waals heterojunction results in a different way of current transport when compared with the traditional lateral p-n diode. The current transport exists in both vertical and lateral directions. Here we try to describe its behavior using a simplified model.

In general, the total resistance of the device is consists of several parts: the contact resistance of metal/$MoS_2$ and metal/black phosphorus, the sheet resistance of $MoS_2$ and black phosphorus, and the resistance of p-n diode at the p-n interface. As shown in Figure S2a, the contact resistance is represented by $R_{cm}$ and $R_{cb}$. The resistances of black hosphorus and MoS$_2$ outside of the overlapped region are represented by R$_b$ and R$_m$, respectively. In the overlapped region, R$_{bo}$ and R$_{mo}$ are the sheet resistances of black phosphorus and MoS$_2$, respectively. The resistance of the diode at the interface is represented by R$_i$. Many diodes are in parallel to represent the p-n interface across the junction region shown in Figure S2a.

For the diode at the interface region, an ideal schematic and band diagrams at 0 bias, forward bias and reverse bias are shown in Figure S2b, c, d. When the diode is in 0 bias, charges generated by incident light are separated at the overlapped region. Under forward bias, the carriers overcome the barrier, so the current increases. Under the reverse bias, the barrier increases which causes the current to decrease. Minority carriers (the electrons/holes in the black phosphorus/MoS$_2$) are extracted to the other side of the junction. As the concentration of minority carriers is quite low, the current is much lower than in forward bias, which generates the current-rectifying IV curves of the diode. Figure S2e shows the band diagram near a contact under various applied voltage biases. Figure S2f shows the band diagram near a contact at different back gate voltage. Figure S2e, f both represent Schottky contacts between metal and MoS$_2$. Similar diagrams can be drawn to represent holes injecting into black phosphorus. Based on these basic models, we discuss the current transport and the related phenomenon in the following sections.

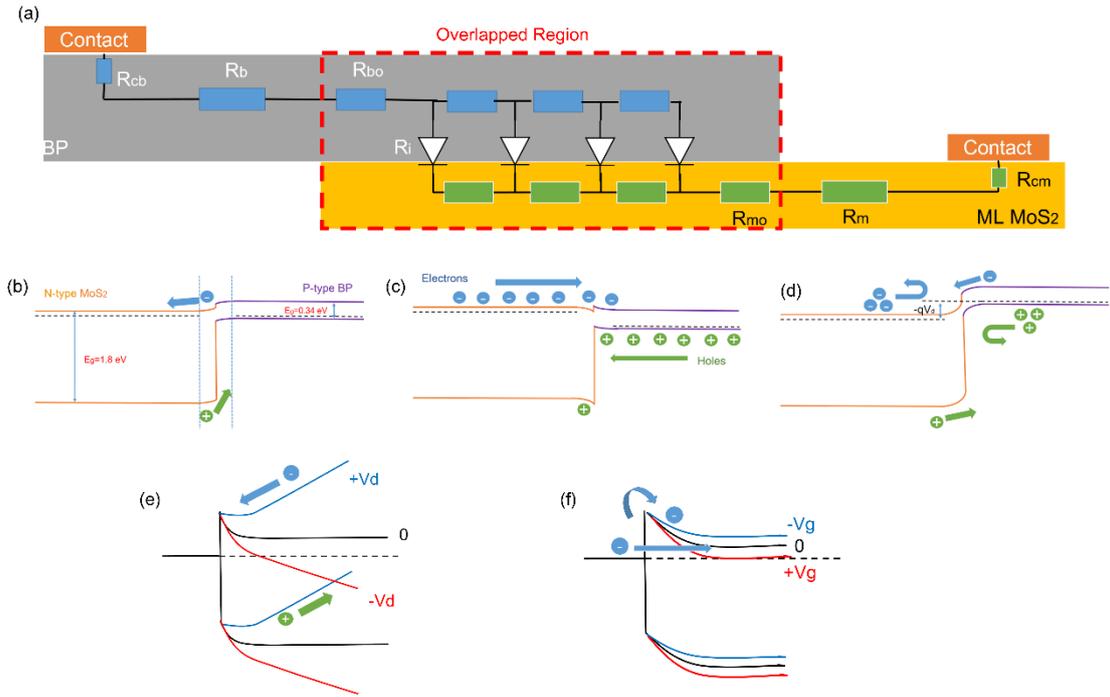

Figure S2 (a) Schematics of the simplified device model. (b) - (d) Schematics of the band diagram at the heterojunction. (e) and (f) Band diagrams near a Schottky contact under various voltage biases (e) and back gate voltages (f).

(1) IV characteristics: The total resistance of the device consists of several parts: p-n interface region, sheet resistance and contact resistance. Here we analyze the effect of these three parts with various back gate voltage to explain the IV characteristics of our p-n diode.

 a) The strong current-rectifying characteristics under forward/reverse bias are by the p-n diode interface region. As shown in the inset (2) of Figure 3a in the main text, this current-rectifying effect can be modulated by the back gate voltage. As the back gate voltage reduces, the rectification ratio increases. This can be explained by the modulation of the band alignment of MoS$_2$ and black phosphorus at the junction region, as shown in Figure S3. Such kind of band alignment is confirmed by the BP experiment in Ref. 1 and $MoS_2$ band alignment calculation, further verifying by this experiment. When a negative/positive back gate voltage is applied, the Fermi levels in both $MoS_2$ and black phosphorus are modulated, as shown in Figure S3a and b. This difference in band alignment results in a different rectification ratio when different back gate voltage is applied.

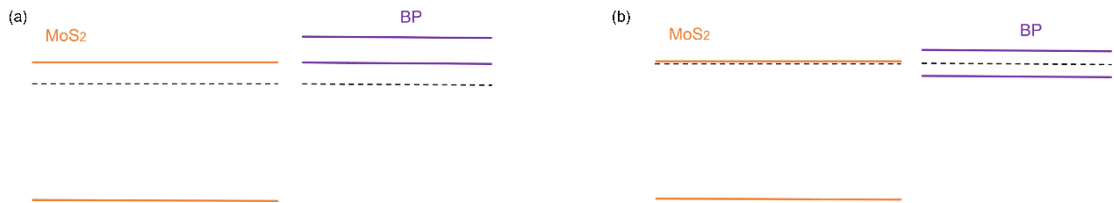

Figure S3 Schematics of the band alignments when (a) negative and (b) positive back gate voltage is applied.

b) The sheet resistance and the contact resistance impact the total current in the device. When increasing the back gate voltage, both the forward and reverse current increase. This is due to these reasons: i) Compared to the few-layer black phosphorus (11 nm in the main text for IV measurements), the sheet resistance of monolayer CVD $MoS_2$ ($R_{mo}$ and $R_m$) is larger.[1-3] So, the total sheet resistance is dominated by the resistance of $MoS_2$. Larger back gate voltages reduce the resistance of $MoS_2$, resulting in a large current. ii) In addition to that, the contact resistance of $MoS_2$/metal can also be reduced as shown in Figure S2f. At the Schottky barrier near the contact, as the back gate voltage increases, the tunneling current can be boosted, though the thermal emission current may not be significantly changed.

(2) Under illumination: Under reverse $V_d$ bias, the band diagram is shown in Figure S2d. The carriers are generated at the p-n interface region under illumination. When increasing the back gate voltage, the sheet/contact resistance is reduced as mentioned above. The band alignments of these two materials at the interface region is also changed as shown in Figure S3. Here, the effect of a positive back gate voltage on reducing the sheet/contact resistance is relatively larger, so the total photocurrent is larger under a positive back gate voltage as shown in Figure 4b. However, when we compare the $I_{illumination}/I_{dark}$ under different back gate voltage, it reaches a maximum value with a $V_g$ of -40 V. If we simplify the p-n interface region as an additional voltage source in this case, then the photocurrent can be expressed as : $I_{illumination}=V_{ph}/R$ where $V_{ph}$ is the voltage source at the p-n interface region and R is the total resistance of the device. The dark current: $I_{dark}=V_d/R$. Thus their ratio can represent the ability of generating electron/hole pairs at the junction region under illumination. The larger $I_{illumination}/I_{dark}$ when the back gate voltage is smaller indicates the band alignment is modulated by the back gate, as shown in Figure S3. This can also explain why the back gate voltage exhibits opposite modulation effect on the short circuit current ($I_{sc}$) and open circuit voltage ($V_{oc}$), as shown in Figure 5c and d.

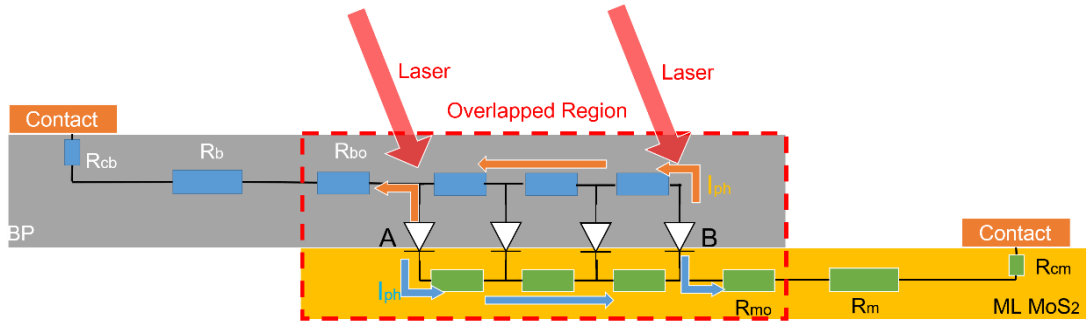

Figure S4 Current flows in the device when different positions (A and B) are under laser illumination during the photocurrent mapping measurement.

(3) Photocurrent mapping: the non-uniform photocurrent generation is due to these reasons: a) non-uniform contact between the $MoS_2$ and black phosphorus due to non-uniform surfaces. b) As we can see in the Figure 6 in the main text, the regions with stronger photocurrent generation are closer to the contact on the $MoS_2$. This is related to the lateral sheet resistance of $MoS_2$ and black phosphorus. Under illumination, the charges are separated at the interface region of the p-n diode. Then they must go through the $MoS_2$ or black phosphorus laterally to reach the metal contact. For simplicity, we treat the diode in Figure S4 at a certain position of the junction region as a voltage source. Here we assume a uniform contact between $MoS_2$ and black phosphorus, which results in a uniform voltage value across the junction region. As shown in Figure S4, when different positions (such as A and B in this figure) of the overlapped region are under laser illumination, the photocurrent flows through the $MoS_2$ and black phosphorus as indicated in the figure. The photocurrent is mainly decided by the lateral sheet resistance of the $MoS_2$ and black phosphorus. Because the sheet resistance of CVD monolayer $MoS_2$ is much larger than the black phosphorus, the resistance of the $MoS_2$ is much more significant. Compared with the positions far from the MoS2/metal contact (such as A in Figure S4), the positions near the MoS2/metal contact (such as B in Figure S4) shows lower $MoS_2$ sheet resistance, resulting in a stronger photocurrent near the contact of $MoS_2$.

3. **Photodetection responsivity of a p-n diode**

The highest photodetection responsivity (R) can be obtained from one particular device. In Figure S5, we can find the maximum R at 1μW is 3.54 A/W for forward bias and 418 mA/W for the reverse bias. It is expected that it could reach a higher value if we use a lower incident laser power for the measurements.

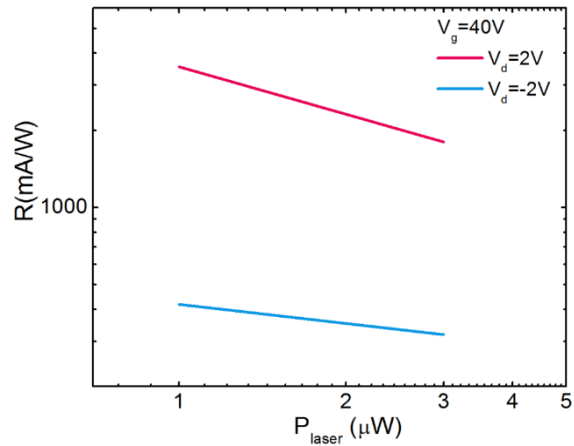

Figure S5 Photodetection responsivity (R) as a function of incident laser power.

4. **The thickness of the flake for the photocurrent mapping**

AFM measurements determined the thickness of the black phosphorus flake used for photocurrent mapping to be about 22 nm, as shown in Figure S6.

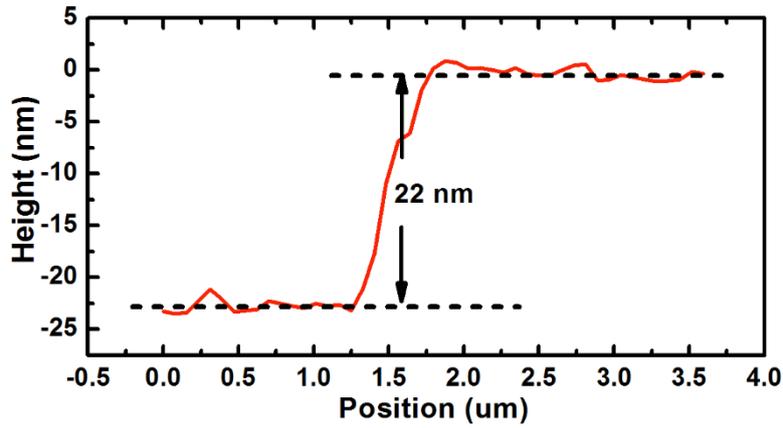

Figure S6 Thickness of the black phosphorus flake for photocurrent mapping.

## 5. Analysis of ideal factor of the p-n diode

In order to understand more of the p-n diode, here we analyze the ideal factor of the fabricated p-n diode in the main text. Two different methods are used to extract the ideal factor.

First, we try to use a simplified circuit model to fit the IV curves of the p-n diode. To achieve a better fitting of curves under both forward and reverse bias, two opposed diodes are connected in series, and each with its own series resistance. It can fits well of both forward and reverse bias region, the fitting curves and the experiment data of the device with zero back gate voltage are shown in Figure S7. Then the ideal factor can be extracted under the forward bias with different back gate voltage, as shown in Figure S8. The ideal factor increases as back gate voltage increases, which can be attributed to the modulation of band alignment with various back gate voltage, as discussed above.

Then, we extract the ideal factor directly from the IV curves using the method from ref. 4 at low voltage regime.[4] The minimum ideal factor is 2.7 with a back gate voltage of -30 V. The relatively large ideal factor can be attributed to large trap density at the interface.

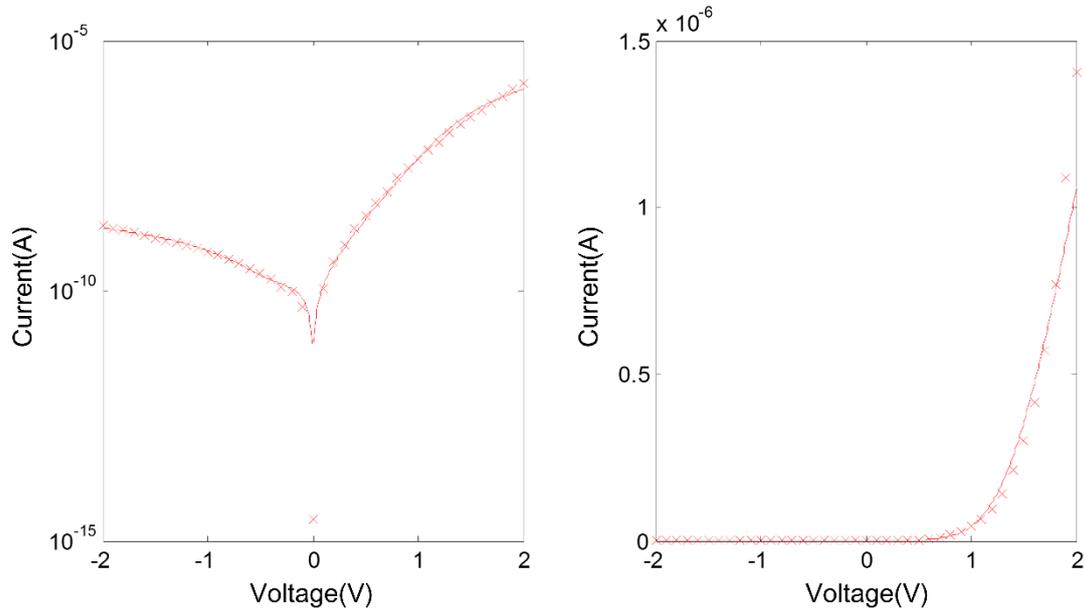

Figure S7 Fitting curves and the experiment data of the p-n diode under semi-log and linear scale.

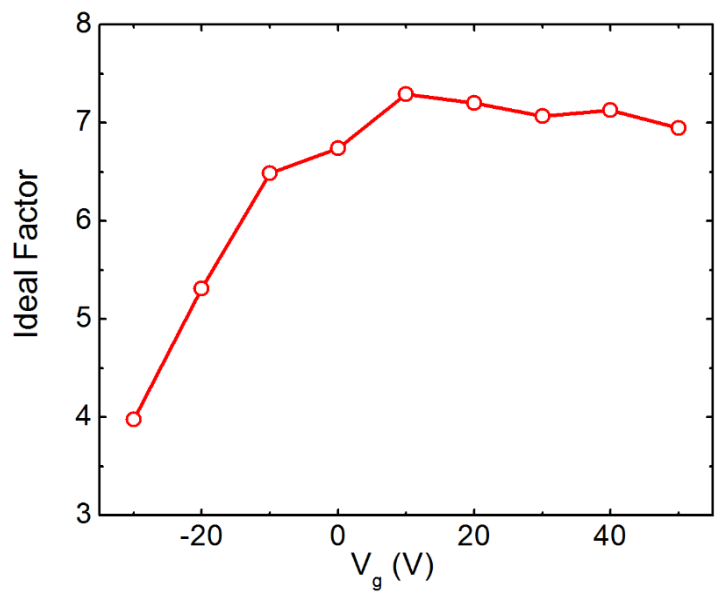

Figure S8 Ideal factor under forward bias as a function of back gate voltage.

References:

(1) Li, L.; Yu, Y.; Ye, G. J.; Ge, Q.; Ou, X.; Wu, H.; Feng, D.; Chen, X. H.; Zhang, Y. *Nat. Nanotechnol.* **2014**.

(2) Liu, H.; Neal, A. T.; Zhu, Z.; Luo, Z.; Xu, X.; Tománek, D.; Ye, P. D. *ACS Nano* **2014**, *8*, 4033-4041.

(3) Liu, H.; Si, M.; Najmaei, S.; Neal, A.; Du, Y.; Ajayan, P.; Lou, J.; Ye. P. D., *Nano Lett.* **2013**, *13*, 2640–2646.

(4) Schroder, D. *Semiconductor Material and Device Characterization*, John Wiley & Sons, **2006**.